\documentclass[useAMS,usenatbib,a4paper]{mn2e}
\usepackage{times}
\usepackage[pdftex]{graphicx}
\newcommand{\B}{\bmath{B}}
\newcommand{\apj}{ApJ}
\newcommand{\mnras}{MNRAS}

\begin{document}

\title[Divergence-free Interpolation]{Divergence-free Interpolation of Vector Fields From Point Values - Exact $\nabla\cdot\bmath{B}=0$  in Numerical Simulations}
\author[Colin P.\ M$^\mathrm{c}$Nally]{Colin P.\ M$^\mathrm{c}$Nally$^{1,2}$\thanks{E-mail:  cmcnally@amnh.org}\\
$^{1}$ Department of Astrophysics, American Museum of Natural History, New York, NY 10024\\
$^{2}$ Department of Astronomy, Columbia University, New York, NY 10027}

\date{Accepted 2011 February 21. Received 2011 February 21; in original form 2011 January 20}
\maketitle

\begin{abstract}
In astrophysical magnetohydrodynamics (MHD) and electrodynamics simulations, numerically enforcing the $\nabla\cdot\bmath{B}=0$  constraint
on the magnetic field has been difficult.
We observe that for point-based discretization, as used in finite-difference type and pseudo-spectral methods,
the $\nabla\cdot\bmath{B}=0$  constraint can be satisfied entirely by a choice of interpolation used to define the derivatives of $\B$.
As an example we demonstrate a new class of finite-difference type derivative operators on a regular grid which has the $\nabla\cdot\bmath{B}=0$  property.
This principle clarifies the nature of $\nabla\cdot\bmath{B}\ne 0$ errors.
The principles and techniques demonstrated in this paper are particularly useful for the magnetic field, but can be applied to any vector field.
This paper serves as a brief introduction to the method and demonstrates an implementation showing convergence.
\end{abstract}


\section{Introduction}
As originally laid out by \citet{1980JCoPh..35..426B}, failing to obey the $\nabla\cdot\bmath{B}=0$  constraint in magnetohydrodynamics
may lead to numerical instability and unphysical results.
This issue has been an issue which has attracted much attention in computational astrophysics
(ex: \citealt{1980JCoPh..35..426B,2004ApJ...602.1079B,2010MNRAS.401.1475P,2009MNRAS.398.1678D}).
To elucidate what $\nabla\cdot\bmath{B}=0$  means, specifying the manner in which $\B$ is represented is essential.
In a numerical method, the vector fields are represented by a discrete set of values.
Two classes of discretizations are popular in astrophysical applications, finite-volume and point values.
Finite volume discretizations store the volume-average of the field over some cell. 
These volume averages constrain the possible divergence of a vector field interpolating these values, and hence
the Constrained Transport method \citep{1988ApJ...332..659E} can be applied to conserve this divergence throughout the simulation.
However, when the magnetic field is represented by point values, the divergence of the interpolated field is not constrained by
the point values, so some extra freedom exists. 
Two classes of approaches have been used. 
The first class is to admit $\nabla\cdot\bmath{B}\ne 0$ errors, and then attempt to manage the consequences.
Methods of this type include the 8-wave scheme \citep{1994arsm.rept.....P,1999JCoPh.154..284P}, and diffusion method \citep{2002JCoPh.175..645D}.
The Smoothed Particle Hydrodynamics schemes of \citet{2004MNRAS.348..123P,2004MNRAS.348..139P,2005MNRAS.364..384P} and \citet{2001ApJ...561...82B,2009MNRAS.398.1678D} also fall into this class,
as the former uses a formulation of the MHD equations which is consistent even in $\nabla\cdot\bmath{B}\ne 0$, and the latter 
removes the $\nabla\cdot\bmath{B}\ne 0$ contributions to the momentum equation.
The second class of methods constrain the derivatives of the interpolated field.
The projection method, used in finite-difference \citep{1980JCoPh..35..426B}, and pseudo-spectral methods,
constructs an interpolation of the magnetic field and then modifies the point values so that with the given interpolation
scheme they produce a divergence-free continuous field.
It is also possible to store and evolve point values of the magnetic vector potential, interpolate this vector 
potential, and find a value and derivatives of the magnetic field from this interpolation. 
This approach is used in the {\sc PENCIL CODE}\footnote{\tt http://pencil-code.googlecode.com/}.
The vector potential approach always yields a magnetic field which is $\nabla\cdot\bmath{B}=0$. 
Some of the disadvantages of this method are that it has the property 
that more than one vector potential configuration leads to the same magnetic field configuration, boundary conditions
may be difficult to arrange, and compared to evolving $\B$ directly an extra level of spatial derivatives needs to be evaluated.

The Smoothed Particle Method (SPH) attempts at MHD are notable in that SPH is a
non-polynomial method used for approximating derivatives. 
In this context, in addition to the aforementioned methods following the strategy of admitting $\nabla\cdot\bmath{B}\ne 0$,
 a method based on the Euler angles formulation have been proposed by \citet{2007MNRAS.377...77P,2007MNRAS.379..915R}
which by construction yields $\nabla\cdot\bmath{B}=0$ , but \citet{2010MNRAS.401..347B} has observed that this
approach is not sufficient for realistic MHD as it severely constrains the allowed magnetic field geometries.
Additionally, 
\citet{2010MNRAS.401.1475P} explored the use of the vector potential strategy in SPH, but found it to be unworkable.

In this paper, we describe a principle that if adhered to allows point-value methods to evolve the magnetic field directly, while maintaining formally $\nabla\cdot\bmath{B}=0$ .
Although throughout this paper we refer to magnetic fields, the principles and methods can apply to any vector field.

\section{A Principle}
Since the discrete point values of the magnetic field do not have a defined derivative, the problem of $\nabla\cdot\bmath{B}=0$  lies 
entirely in the method used to produce the continuous representation of the magnetic field from which derivatives are
taken.
Thus, to produce a $\nabla\cdot\bmath{B}=0$  method, it is sufficient to define an interpolation (or quasi-interpolation) 
which is restricted to producing only $\nabla\cdot\bmath{B}=0$  fields.
A concrete example of such a method is provided by divergence-free matrix valued radial basis function interpolation
 \citep{narcowichandward,LowitzschPhD,Lowitzsch2005}.
The following two sections of this paper are devoted to a summary of this technique, and its use to construct finite-difference like operators.

Radial Basis Function (RBF) Interpolation is an alternate method to polynomial basis interpolation for constructing
functions which interpolate some discrete set of values.
Instead of using a set of functions with a different polynomial form all mentored at the same place (a Taylor Series),
RBF uses shifted versions of a one-parameter function.
Further, these functions are shifted to be centred on each interpolating point.
For a set of scalar valued samples 
$\{\bmath{x}_j,d_j\}^N_{j=1}$ where $\bmath{x}_j$ is the position of each point and $d_j$ 
is the value of the scalar field to be interpolated at that point,  the RBF interpolant is of the form
\begin{equation}
s(x) = \sum_{j=1}^{N} \psi(\|\bmath{x}-\bmath{x}_j\|)c_j
\label{rbf}
\end{equation}
where $s(x)$ is the interpolant, $\psi$ is a radial basis function, and $c_j$ are a set of coefficients.
These coefficients $\{c_j\}_{j=1}^N$ are such that 
\begin{equation}
s(x_k) = d_k \mathrm{\ for\ all\ } 1 \leq k \leq N.
\end{equation}
Solving for these coefficients is done by solving
the equation $\bmath{G}\bmath{c} = \bmath{d}$
where the matrix entries $\bmath{G}_{i,j} = \Psi(\|x_i-x_j\|)$.
The remarkable ability of RBF interpolation is that if $\psi$ has certain properties,
this equation has a unique solution for any set of points $\{x_j\}$ in any number of dimensions.
The reader is encouraged to refer to  \citet{wendland} and \citet{buhmann} for the mathematical details of the theory of radial basis function 
interpolation.

Beyond scalar fields, it is possible to construct a RBF interpolation for a vector field such as the magnetic field.
If the RBF is chosen appropriately, this interpolation can be constrained to produce $\nabla\cdot\bmath{B}=0$.
Given a set of point values $\{\bmath{x}_j,\bmath{d}_j\}^N_{j=1}$ where $\bmath{x}_j$ is the position of each point and $\bmath{d}_j$ 
is the value of the vector field to be interpolated at that point, the interpolation is constructed in the form
\begin{equation}
\bmath{s}(\bmath{x}) = \sum_{j=1}^{N} \Phi(\|\bmath{x}-\bmath{x}_j\|)\bmath{c}_j
\label{mrbf}
\end{equation}
where $\{\bmath{c}_j\}_{j=1}^N$ are such that 
\begin{equation}
\bmath{s}(\bmath{x}_k) = \bmath{d}_k \mathrm{\ for\ all\ } 1 \leq k \leq N
\end{equation}
The matrix valued radial basis function $\Phi$ is constructed by
\begin{equation}
\Phi(\bmath{x}) = \{\nabla \nabla^T - \nabla^2\bmath{I}\}\psi(\bmath{x})
\label{phieq}
\end{equation}
where $\psi$ is a scalar valued radial basis function and $\bmath{I}$ is an identity matrix.
If a numerical method  is built using this representation for the magnetic fields, then the results will be 
free of $\nabla\cdot\bmath{B}\ne 0$ errors.
This use of a $\nabla\cdot\bmath{B}=0$  interpolation basis is a general principle, it could apply to other classes of basis, and spectral basis.

\section{Demonstration}

\begin{figure}
\includegraphics[width=8cm]{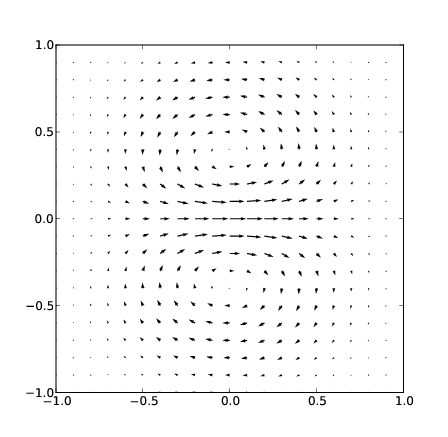}
\caption{The $x$ vector field component of eq.\ \ref{Phiform}  with $\epsilon=3.5$, the $y$ component is the same rotated $90$ degrees.}
\label{rbfbasisfields}
\end{figure}

\begin{figure}
\includegraphics[width=9cm]{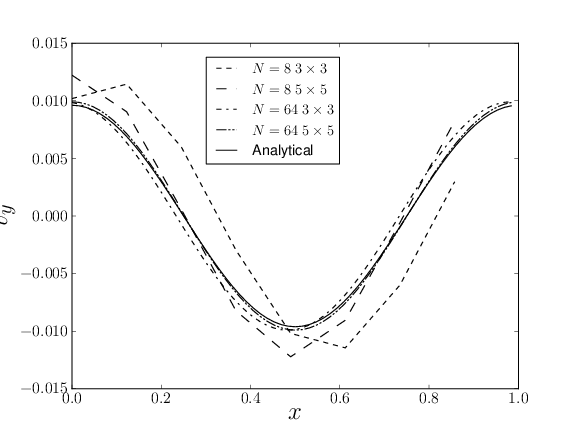}\\
\includegraphics[width=9cm]{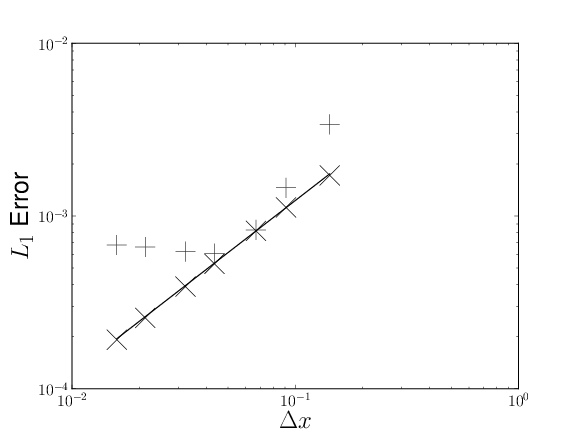}
\caption{{\sl Upper}: Alfven wave solutions at two resolutions, grid size $N\times N$ with $3\times 3$ and $5 \times 5$ point stencils. {\sl Lower}: Convergence of Alfven wave solution to analytical result. Results from $3\times 3$ stencils plotted with $+$ and from $5 \times 5$ stencils plotted with $\times$. 
The solid line marks a slope of $1$.
The $3\times 3$ stencil error saturates at a larger value than the $5\times 5$ stencil error.}
\label{alfvenconvergence}
\end{figure}

In a  manner similar to Taylor-series based finite difference stencils, we can build
generalised finite difference stencils using radial basis function interpolation. 
The procedure is the same as for Taylor-series based finite differences, we interpolate the data at a local set of points,
then take derivatives of the interpolant.
Like with Taylor-series based finite differences, the resulting scheme will not in general conserve
mass, linear momentum, or energy to machine precision.
These quantities will usually only be conserved to the level of the truncation error of the scheme.
One can look to the body of work produced with the {\sc PENCIL CODE}, a high order finite difference scheme, to
see examples of a successful approach based on a non-conservative method (ex: \citet{2007Natur.448.1022J,2011JCoPh.230....1B}).
Scalar value radial basis function finite difference stencils have been studied in \citet{2010JCoPh.229.8281B} 
for the case of the multiquadric radial basis function. The radial basis function finite difference approach (RBF-FD)
has also been applied to convection-type PDEs in \citet{Fornberg2010}.

To illustrate this construction, we must choose a radial basis function, in this case  a Gaussian:
\begin{equation}
\psi(r) = e^{-\epsilon r^2}
\end{equation}
where $\epsilon$ is a constant called the scaling factor. 
The scaling factor can be adjusted depending on the interpolation point distribution.
Other radial basis functions can be used (see \citealt{wendland} or \citealt{buhmann}, and recent results on the near equivalence of 
some common RBFs \citealt{boyd}), but the Gaussian gives the simplest algebraic expressions in the following.

To construct a divergence free matrix valued basis function from $\psi(r)$, we apply equation \ref{phieq}
in two dimensions with $r^2 = x^2+y^2$, yielding:
\begin{eqnarray}
\Phi_{11} &=& -(4\epsilon^2 y^2 -2 \epsilon)e^{-\epsilon (x^2+y^2)}\nonumber\\
\Phi_{12} &=& 4\epsilon^2 e^{-\epsilon (x^2+y^2)} \label{Phiform}\\
\Phi_{21} &=& \Phi_{12}\nonumber\\
\Phi_{22} &=& -(4\epsilon^2 x^2 -2 \epsilon)e^{-\epsilon (x^2+y^2)}\nonumber
\end{eqnarray}
The combinations $(\Phi_{11},\Phi_{12})$ and  $(\Phi_{21},\Phi_{22})$ are divergence-free vector fields.
Figure \ref{rbfbasisfields} shows the two components. One component resembles a dipole field in the $x$ direction,
and the other a dipole field in the $y$ direction.
In essence the method interpolates only $\nabla\cdot\bmath{B}=0$  fields because the field is built entirely
from shifted and normalised versions of these dipole components.
To build up an interpolation of some point-sampled field with these as the basis functions, it is necessary to solve
the set of linear equations:
\begin{equation}
\bmath{A}\bmath{c}=\bmath{d}
\end{equation}
For  $N$ interpolation points the matrix $A$ has entries:
\begin{eqnarray}
\bmath{A}_{1:N,1:N} &\rightarrow& \bmath{A}_{i,j} = \Phi_{11}(r_{ij})\\
\bmath{A}_{N+1:2N,1:N}  &\rightarrow& \bmath{A}_{i,j} = \Phi_{12}(r_{ij})\\
\bmath{A}_{1:N,N+1:2N}  &\rightarrow& \bmath{A}_{i,j} = \Phi_{21}(r_{ij})\\
\bmath{A}_{N+1:2N,N+1:2N}  &\rightarrow& \bmath{A}_{i,j} = \Phi_{22}(r_{ij})
\end{eqnarray}
Each sub-matrix of $\bmath{A}$ has entries corresponding to an entry in $\Phi$.
The vector $\bmath{d}$ has entries:
\begin{eqnarray}
\bmath{d}_{1:N} &\rightarrow& \bmath{d}_i = B_{i,x} -B_{x0}\\
\bmath{d}_{N+1:2N} &\rightarrow& \bmath{d}_i = B_{i,y} -B_{y0}
\end{eqnarray}
$\bmath{A}$ is the interpolation matrix, and $\bmath{d}$ is the values being interpolated.
$B_{i,x}$ and $B_{i,y}$ are the components of the vector field being interpolated.
$B_{x0}$ and $B_{y0}$ are constant background field components, which may be chosen
to be the field at the interpolation point where the derivatives are being calculated.
This subtraction of the background constant component of the field increases the accuracy of the radial basis function
approximation as this component is not in the space spanned by the interpolation basis.
This background component is irrelevant to the $\nabla\cdot\bmath{B}=0$  constraint and to the determination of derivatives.
The vector $\bmath{c}$ is composed of the interpolation coefficients in eq. \ref{mrbf}.
To find the derivatives of the interpolating function at point $\bmath{x}_0$, we can simply evaluate the derivative of eq.\ \ref{mrbf} as
\begin{equation}
\frac{\partial \bmath{s}}{\partial x}\Big |_{\bmath{x}=\bmath{x}_0} = \sum_{j=1}^{N} \frac{\partial \Phi}{\partial x}\Big |_{\bmath{x}=\bmath{x}_0-\bmath{x}_j}\bmath{c}_j
\end{equation}
which yields the radial basis function estimate of the derivative at the point $\bmath{x}_0$.
This gives us a method of finding the derivatives of a $\nabla\cdot\bmath{B}=0$  magnetic field from point values.
The interpolation points chosen can be arbitrary, but for the purposes of building finite-difference like derivative operators
a set of nearest neighbouring points is natural choice.
In the following, we demonstrate the use of $3\times 3$ (9 point) and $5\times 5$ (25 point)  stencils, centred on $\bmath{x}_0$, in two dimensions to solve the equations of 
magnetohydrodynamics.

The equations solved are those for viscous, resistive, compressible isothermal magnetohydrodynamics 
in two dimensions:
\begin{eqnarray}
\frac{\partial \rho}{\partial t} &=& -\nabla \cdot (\rho\bmath{v})\\
\frac{\partial \rho\bmath{v}}{\partial t} &=& \bmath{v}\cdot\nabla(\rho\bmath{v}) -\nabla P +(\nabla\times\B)\times \B + \nu\nabla^2(\rho\bmath{v})\\
\frac{\partial \B}{\partial t} &=& \nabla \times (\bmath{v}\times\B) + \eta\nabla^2\B
\end{eqnarray}
with the equation of state $P  = c_s^2 \rho$
where $\rho$ is the density, $\bmath{v}$ is the velocity, $P$ is pressure, $\B$ is the magnetic field, $\nu$
 is the dynamic viscosity, and $\eta$ is the magnetic diffusivity.
The equations are spatially discretized on an evenly spaced square grid with periodic boundary conditions.
Spatial derivatives are estimated using the radial basis function methods on $3\times 3$ stencils as outlined above,
 and explicit time integration is performed with the forward Euler method.
Both the derivatives of the scalar fields ($\rho$,$P$, components of $\bmath{v}$) 
and the vector field $\B$ are taken with scalar and vector RBF interpolations.
This method is chosen so that the resulting code is as simple as possible to facilitate the reader's understanding.
The source code in Python is available on the author's website\footnote{ {\tt http://www.astro.columbia.edu/$\sim$colinm/dfi/}}.

\begin{figure*}
\begin{center}
\includegraphics[width=18cm]{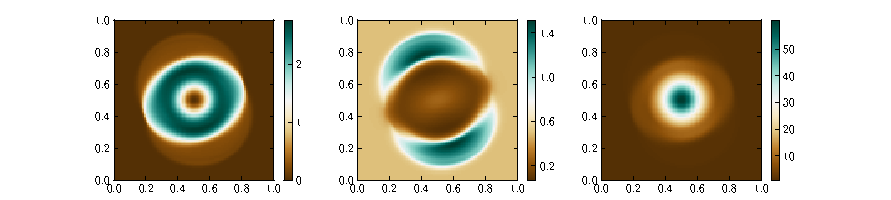}
\caption{Magnetised blast problem {\sl Left}: Kinetic energy density {\sl Middle}: Magnetic energy density {\sl Right}: Mass density}
\end{center}
\label{rbfblast}
\end{figure*}

\begin{figure}
\begin{center}
\includegraphics[width=9cm]{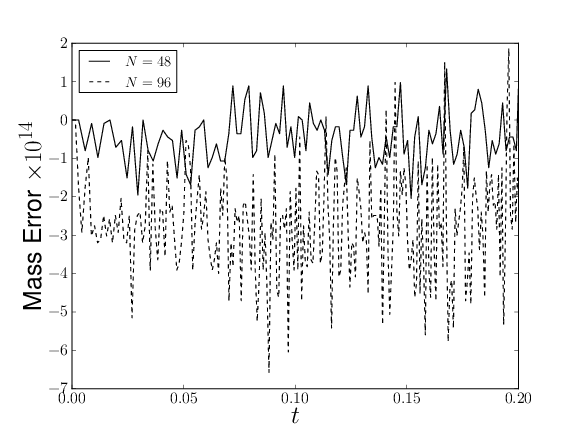}
\caption{Magnetised blast problem total mass error for two grid resolutions $N=48$ and $N=96$}
\label{rbfblastmass}
\end{center}
\end{figure}

\begin{figure}
\begin{center}
\includegraphics[width=9cm]{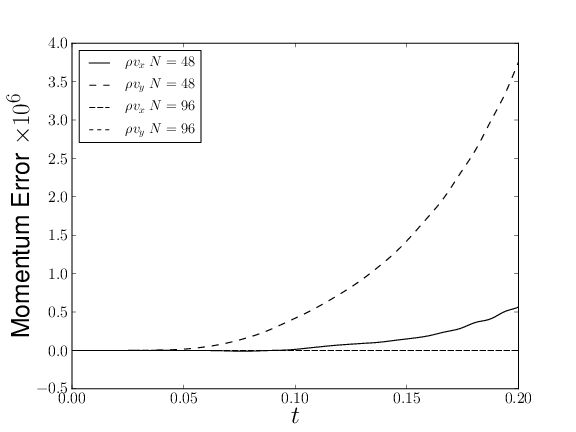}
\caption{Magnetised blast problem momentum errors for two grid resolutions $N=48$ and $N=96$}
\label{rbfblastmomentum}
\end{center}
\end{figure}

\begin{table}
\begin{center}
\caption{Linear momentum errors in the magnetised blast problem}
\label{blasterrors}
\begin{tabular}{|l|llll}
\hline
Error & $N=32$ &$N=48$ &$N=64$ &$N=96$ \\
\hline
$\rho v_x$ & $5.6\times10^{-6}$ & $5.7\times10^{-7}$ & $1.2\times10^{-8}$ & $1.5\times10^{-12}$\\
$\rho v_y$ & $1.5\times10^{-4}$ & $3.8\times10^{-6}$ & $5.9\times10^{-8}$ & $3.0\times10^{-17}$ \\
\hline
\end{tabular}
\end{center}
\end{table}

Since the method is $\nabla\cdot\bmath{B}=0$  by construction, we need only demonstrate that the solution converges as $\nabla\cdot\bmath{B}\ne 0$ errors cannot occur.
A suitable test is the evolution of a damped Alfven wave for finite $\nu$ and $\eta$, the analytical
solution for which is given in \citet{chandrasekhar}, section 39.
The experimental convergence of the numerical solution to the analytical result for an
 Alfven  wave is shown in Figure 
\ref{alfvenconvergence} with $\epsilon=1/64$, $\nu=\mu=0.001$. 
Note that the error saturates in this test for the $3\times3$ stencil.
As the RBF interpolation used does not reproduce a first-order polynomial exactly, the approximation effectively stops
improving below some grid spacing. 
To obtain further convergence, a larger stencil must be used.
The $L_1$ error when the $5\times 5$ point stencil is used decays  at a rate of $1.0$, which is the limit set by the first order time integration scheme.

As a further demonstration, the result of a magnetised blast problem, starting for an initial over density in the centre of the box
is shown in Figure \ref{rbfblast}.
The initial condition of this problem is $\nu =\eta=0.005$, $c_s=0.4082$, $\rho = 99e^{-((x-0.5)^2+(y-0.5)^2/0.12^2)^2}+1 $, $\bmath{v}=0$,
$B_x=\cos(2\pi/21)$,  $B_y=\sin(2\pi/21)$ 
in a area $1\times 1$ with a 
$96 \times 96$ grid with periodic boundary conditions.
The output is shown at $t=0.2$.
The $3\times3$ stencil was used with $\epsilon=1/16$.
A similar problem is shown in \citet{2008ApJS..178..137S}.
We note the the intermediate shock can be seen in the solution on the axis of the blast along the magnetic field \citep{1991ApJ...375..239F}.
The time history of the absolute error in total mass in the magnetised blast problem is shown in Figure \ref{rbfblastmass} for two resolutions. 
The error is at the limit of numerical precision for all resolutions. Evidently the scheme used here is in fact mass-conserving even though it was not 
explicitly constructed to be so.
The error in linear momentum is limited by truncation error, Figure \ref{rbfblastmomentum} shows the time evolution of the absolute momentum error for two resolutions.
Table \ref{blasterrors} lists the momentum errors at $t=0.2$ in the blast problem for four grid resolutions $N$.
The momentum error can be seen to converge towards zero as the resolution is increased.
Energy conservation errors are not treated as the discussion is limited to isothermal magnetohydrodynamics.

The stability of this scheme is not explored here as the method is presented only as a demonstration of the use of these 
radial basis function based derivative operators. 
The computational cost of a RBF based finite difference scheme for a derivative is the same as
that for a polynomial based scheme with the same number of stencil points, as the only difference
is in the stencil coefficients.
However, with RBF based schemes the well-motivated stencils may not be the same shape and size as for polynomial
based finite differences - for example the square $3\times3$ stencil used here is not a popular choice when used with
polynomial based schemes.
In contrast to the most directly comparable polynomial based finite difference scheme yielding $\nabla\cdot\bmath{B}=0$  ( the 
vector potential method), the RBF based approach has the advantage of requiring fewer derivative stencils to be computed
as the magnetic field is obtained directly not computed from derivatives of the vector potential.

\section{Extensions}
The method for constructing radial basis function based derivative approximations
 has no dependence on regularly placed points or the existence of mesh edges.
Hence, these methods are easily used in a mesh-free context. 
Also, though in radial basis function interpolation theory the interpolation points are chosen
to be the radial basis function centres, in practise the approximation matrix can still be inverted
if the interpolation points do not coincide with the radial basis function centres.
Furthermore, fewer radial basis functions can be used than interpolation points are specified - in this case
a $\nabla\cdot\bmath{B}=0$  least squares approximation can be computed.
The divergence-free basis used as an example in this work is not orthonormal.
If an orthonormal basis were specified, it would be possible to preform divergence-free
pseudo-spectral simulations with $\B$, which may be of particular use in general relativistic MHD and 
force-free electrodynamics simulations.

Other constraints beyond divergence-free can be placed on the vector field. For example, \citet{Lowitzsch2005}
observed that $\nabla \times \B = 0$ type vector fields can be interpolated in a similar manner to shown here.
This suggests the possibility to satisfy more complicated, though homogeneous, constraints.

\section{Conclusion}
In a point-value method, $\nabla\cdot\bmath{B}=0$  can be satisfied by the correct choice of interpolation scheme.
Matrix-valued radial basis functions provide such an interpolations scheme.
Finite-difference-like $\nabla\cdot\bmath{B}=0$  derivative operators can be constructed from matrix-valued radial basis function
interpolations, and their use in the solution of magnetohydrodynamics problems has been demonstrated.
Further exploration of the stability properties, accuracy, and computational cost of schemes based on these operators is warranted. 
The underlying principle of the choice of a $\nabla\cdot\bmath{B}=0$  interpolation also applies to pseudo spectral methods, and
in general can be applied to any vector field where such a constraint is required.

\section{Acknowledgements}
We acknowledge useful discussions with M.-M.~Mac~Low and J.~Maron for discussions on the nature of $\nabla\cdot\bmath{B}=0$  and  
M.-M.~Mac~Low for advice on the manuscript. C.P.M.\ was supported by NSF CDI grant AST08-35734.


\begin{thebibliography}{28}
\expandafter\ifx\csname natexlab\endcsname\relax\def\natexlab#1{#1}\fi

\bibitem[{{Babkovskaia}, {Haugen} \& {Brandenburg}(2011){Babkovskaia},
  {Haugen}, \& {Brandenburg}}]{2011JCoPh.230....1B}
{Babkovskaia} N., {Haugen} N.~E.~L., {Brandenburg} A., 2011, Journal of
  Computational Physics, 230, 1

\bibitem[{{Balsara} \& {Kim}(2004)}]{2004ApJ...602.1079B}
{Balsara} D.~S., {Kim} J., 2004, \apj, 602, 1079

\bibitem[{{Bayona} {et~al}\mbox{.}(2010){Bayona}, {Moscoso}, {Carretero}, \&
  {Kindelan}}]{2010JCoPh.229.8281B}
{Bayona} V., {Moscoso} M., {Carretero} M., {Kindelan} M., 2010, Journal of
  Computational Physics, 229, 8281

\bibitem[{{B{\o}rve}, {Omang} \& {Trulsen}(2001){B{\o}rve}, {Omang}, \&
  {Trulsen}}]{2001ApJ...561...82B}
{B{\o}rve} S., {Omang} M., {Trulsen} J., 2001, \apj, 561, 82

\bibitem[{{Boyd}(2010)}]{boyd}
{Boyd} J.~P., 2010, Journal of Computational Physics

\bibitem[{{Brackbill} \& {Barnes}(1980)}]{1980JCoPh..35..426B}
{Brackbill} J.~U., {Barnes} D.~C., 1980, Journal of Computational Physics, 35,
  426

\bibitem[{{Brandenburg}(2010)}]{2010MNRAS.401..347B}
{Brandenburg} A., 2010, \mnras, 401, 347

\bibitem[{{Buhmann}(2003)}]{buhmann}
{Buhmann} M.~D., 2003, {Radial Basis Functions}. {Cambridge}

\bibitem[{{Chandrasekhar}(1961)}]{chandrasekhar}
{Chandrasekhar} S., 1961, {Hydrodynamic and hydromagnetic stability}. Oxford:
  Clarendon

\bibitem[{{Dedner} {et~al}\mbox{.}(2002){Dedner}, {Kemm}, {Kroner}, {Munz},
  {Schnitzer}, \& {Wesenberg}}]{2002JCoPh.175..645D}
{Dedner} A., {Kemm} F., {Kroner} D.~., {Munz} C., {Schnitzer} T., {Wesenberg}
  M., 2002, Journal of Computational Physics, 175, 645

\bibitem[{{Dolag} \& {Stasyszyn}(2009)}]{2009MNRAS.398.1678D}
{Dolag} K., {Stasyszyn} F., 2009, \mnras, 398, 1678

\bibitem[{{Evans} \& {Hawley}(1988)}]{1988ApJ...332..659E}
{Evans} C.~R., {Hawley} J.~F., 1988, \apj, 332, 659

\bibitem[{{Ferriere}, {Mac Low} \& {Zweibel}(1991){Ferriere}, {Mac Low}, \&
  {Zweibel}}]{1991ApJ...375..239F}
{Ferriere} K.~M., {Mac Low} M., {Zweibel} E.~G., 1991, \apj, 375, 239

\bibitem[{Fornberg \& Lehto(2010)}]{Fornberg2010}
Fornberg B., Lehto E., 2010, Journal of Computational Physics, In Press,
  Accepted Manuscript,

\bibitem[{{Johansen} {et~al}\mbox{.}(2007){Johansen}, {Oishi}, {Mac Low},
  {Klahr}, {Henning}, \& {Youdin}}]{2007Natur.448.1022J}
{Johansen} A., {Oishi} J.~S., {Mac Low} M., {Klahr} H., {Henning} T., {Youdin}
  A., 2007, Nature, 448, 1022

\bibitem[{{Lowitzsch}(2002)}]{LowitzschPhD}
{Lowitzsch} S., 2002, PhD thesis, {Texas A\&M University}

\bibitem[{{Lowitzsch}(2005)}]{Lowitzsch2005}
---, 2005, Journal of Approximation Theory, 137, 238

\bibitem[{{Narcowich} \& {Ward}(1994)}]{narcowichandward}
{Narcowich} F., {Ward} J., 1994, Mathematics of Computation, 63, 661

\bibitem[{{Powell}(1994)}]{1994arsm.rept.....P}
{Powell} K.~G., 1994, {Approximate Riemann solver for magnetohydrodynamics
  (that works in more than one dimension)}. Tech. Rep. ICASE Report No. 94-24,
  Institute for Computer Applications in Engineering and Science, NASA Langley
  Research Center

\bibitem[{{Powell} {et~al}\mbox{.}(1999){Powell}, {Roe}, {Linde}, {Gombosi}, \&
  {de Zeeuw}}]{1999JCoPh.154..284P}
{Powell} K.~G., {Roe} P.~L., {Linde} T.~J., {Gombosi} T.~I., {de Zeeuw} D.~L.,
  1999, Journal of Computational Physics, 154, 284

\bibitem[{{Price}(2010)}]{2010MNRAS.401.1475P}
{Price} D.~J., 2010, \mnras, 401, 1475

\bibitem[{{Price} \& {Bate}(2007)}]{2007MNRAS.377...77P}
{Price} D.~J., {Bate} M.~R., 2007, \mnras, 377, 77

\bibitem[{{Price} \& {Monaghan}(2004{\natexlab{a}})}]{2004MNRAS.348..123P}
{Price} D.~J., {Monaghan} J.~J., 2004{\natexlab{a}}, \mnras, 348, 123

\bibitem[{{Price} \& {Monaghan}(2004{\natexlab{b}})}]{2004MNRAS.348..139P}
---, 2004{\natexlab{b}}, \mnras, 348, 139

\bibitem[{{Price} \& {Monaghan}(2005)}]{2005MNRAS.364..384P}
---, 2005, \mnras, 364, 384

\bibitem[{{Rosswog} \& {Price}(2007)}]{2007MNRAS.379..915R}
{Rosswog} S., {Price} D., 2007, \mnras, 379, 915

\bibitem[{{Stone} {et~al}\mbox{.}(2008){Stone}, {Gardiner}, {Teuben}, {Hawley},
  \& {Simon}}]{2008ApJS..178..137S}
{Stone} J.~M., {Gardiner} T.~A., {Teuben} P., {Hawley} J.~F., {Simon} J.~B.,
  2008, ApJS, 178, 137

\bibitem[{{Wendland}(2005)}]{wendland}
{Wendland} H., 2005, {Scattered Data Approximation}. {Cambridge}

\end{thebibliography}

\end{document}